\RequirePackage[hyphens]{url}
\documentclass[journal=jpccck,manuscript=article]{achemso}

\usepackage[version=3]{mhchem} 
\usepackage{xcolor}
\usepackage{soul} 
\usepackage{amssymb}
\usepackage{lineno,hyperref}
\usepackage{graphicx}
\usepackage{dcolumn}
\usepackage{bm}
\usepackage{float}
\usepackage{lipsum}

\newcommand{\DIPC}[0]{
Donostia International Physics Center (DIPC),
Paseo Manuel de Lardizabal 4, 20018 Donostia-San Sebasti\'an, Spain}
\newcommand{\CFM}[0]{
Centro de F\'{\i}sica de Materiales CFM/MPC (CSIC-UPV/EHU), Paseo Manuel de Lardizabal 5, 20018 Donostia-San Sebasti\'an, Spain}
\author{Auguste TETENOIRE}
\email{auguste.tetenoire@dipc.org}
 \affiliation{\DIPC}

\author{J.\ I.\ Juaristi}
\email{josebainaki.juaristi@ehu.eus}
\affiliation{Departamento de Pol\'{\i}meros y Materiales Avanzados: F\'{\i}sica, Qu\'{\i}mica y Tecnolog\'{\i}a, Facultad de Qu\'{\i}micas (UPV/EHU), Apartado 1072, 20080 Donostia-San Sebasti\'an, Spain}
\alsoaffiliation{\CFM}
\alsoaffiliation{\DIPC}

\author{M. Alducin}
\email{maite.alducin@ehu.eus}
 \affiliation{\CFM}
 \alsoaffiliation{\DIPC}

\title{Disentangling the Role of Electrons and Phonons in the Photoinduced CO Desorption and CO Oxidation on (O,CO)-Ru(0001)}

\date{\today}

\begin{document}

\begin{abstract}\label{sec:abstract}
The role played by electronic and phononic excitations in the femtosecond laser induced desorption and oxidation of CO coadsorbed with O on Ru(0001) is investigated using ab initio molecular dynamics with electronic friction. To this aim, simulations that account for both kind of excitations and that only consider electronic excitations are performed. Results for three different surface coverages are obtained. We unequivocally demonstrate that CO desorption is governed by phononic excitations. In the case of oxidation the low statistics does not allow to give a categorical answer. However, the analysis of the adsorbates kinetic energy gain and displacements strongly suggest that phononic excitations and surface distortion also play an important role in the oxidation process.
\end{abstract}

\section{Introduction}

The use of intense femtosecond laser pulses in the near infrared, visible, and ultraviolet regime constitutes an efficient tool to promote adsorbate reactions at metal surfaces that are forbidden or less likely under thermal conditions~\citep{cavanagh93,guo99,frischkorncr06,saalfrankcr06}. The laser excites the electrons of the metal and energy is subsequently transferred to the surface atoms by means of electron-phonon coupling. As a consequence, the adsorbates can gain energy from both the excited electronic and phononic systems. Experimentally, two-pulse correlation measurements have been used to disentangle which the timescale for the energy transfer to the adsorbates is~\citep{frischkorncr06,budde91,busch95,bonn99,funk00,denzler03,sung16,szymanskijpc07}. In this way, the reaction is ascribed to be a dominant electron-assisted process when its timescale is of few picoseconds or less and to be a dominant phonon-assisted process when its timescale is longer.

From the theoretical side, a proper understanding of this kind of experiments and of their outcome requires a proper characterization of the reaction dynamics. The excitation generated by the laser on the surface is accounted for using a two temperature model (2TM) in which the electronic and phononic excitations are described in terms of time-dependent electronic ($T_e$) and phononic ($T_l$) temperatures~\citep{anisimov74}. Subsequently, the dynamics of the adsorbates in the highly excited environment are simulated~\citep{tullyss94,springer96,vazha09, fuchsel10,fuchsel11,loncaricprb16,loncaricnimb16,scholz16}. In this respect, the extension of the ab initio molecular dynamics with electronic friction method~\citep{novkoprb15,novkoprb16,novkoprb17} to incorporate the effect of time-dependent electronic and phononic temperatures in the adsorbate dynamics [hereafter denoted as ($T_\textrm{e}, T_\textrm{l}$)-AIMDEF]~\citep{alducin2019,scholz19,tetenoire2022,tetenoire2023} constitutes a way of treating the multidimensional dynamics of the adsorbates and surface atoms at the density functional theory (DFT) level, incorporating the coupling of the adsorbates to both the excited electronic and phononic systems. 

An important reaction that cannot be thermally activated under ultrahigh vacuum conditions~\citep{Kostov1992Nov,bonn99} but can be propelled by femtosecond laser pulses~\citep{bonn99,obergJCP2015,ostrom2015Feb} is CO oxidation when coadsorbed with atomic O on the Ru(0001) surface. Still, even in these conditions, CO desorption is around 30 times more probable than CO oxidation~\citep{bonn99,obergJCP2015}. In two previous works~\citep{tetenoire2022,tetenoire2023}, we have applied the ($T_\textrm{e}, T_\textrm{l}$)-AIMDEF method to this system. Different surface coverages, for which the reaction paths under equilibrium conditions for CO desorption and oxidation had been previously studied~\citep{tetenoire2021}, were taken into account. Our results reproduced the experimental fact regarding the CO desorption to oxidation branching ratio being larger than one order of magnitude. Additionally, our dynamics simulations showed the reason for this behavior. We observed that CO desorption is a direct process only limited by the energy the CO molecules need to gain to overcome the desorption energy barrier. In contrasts, the oxidation dynamics is much more complex, the configurational space to oxidation is very restricted, and the fact that the O and CO adsorbates gain energy enough to overcome the energy barrier to oxidation does not guarantee their recombination. Our simulations also reproduced the changes in the O K-edge XAS experimental spectra attributed to the initial stage of the oxidation process~\citep{ostrom2015Feb}, further confirming the robustness of the theoretical model.

An important question that was not studied in the previous works is the relative importance of electronic and phononic excitations in both CO desorption and CO oxidation reactions. In the present paper we aim to elucidate this question. 
We perform the so-called $T_\textrm{e}$-AIMDEF simulations~\citep{juaristiprb17}, in which the Ru surface atoms are kept frozen in their equilibrium positions, so that the adsorbates are uniquely coupled to the excited electrons. In this way, we gain information about the CO desorption and oxidation probabilities, and about the dynamics of these processes, when only electronic excitations are considered. Comparison of these results with those obtained in the ($T_\textrm{e}, T_\textrm{l}$)-AIMDEF simulations, in which the effect of both electronic and phononic excitations is accounted for, allows us to answer the question about which channel dominates each reaction on each of the studied surface coverages.

The paper is organized as follows. The theoretical model and computational settings are described in the Theoretical Methods section. The results of both the $T_\textrm{e}$-AIMDEF and the ($T_\textrm{e}, T_\textrm{l}$)-AIMDEF simulations for the CO desorption and oxidation probabilities, kinetic energy gains, and adsorbate displacements are presented in the Results and Discussions section. Finally, the main conclusions of the paper are summarized in the Conclusions section. 


\section{Theoretical Methods \label{sec:methods}}
\subsection{Photoinduced Desorption Model\label{sec:theory}}

\begin{figure}[!hb]
\begin{center}
\includegraphics[width=1\columnwidth]{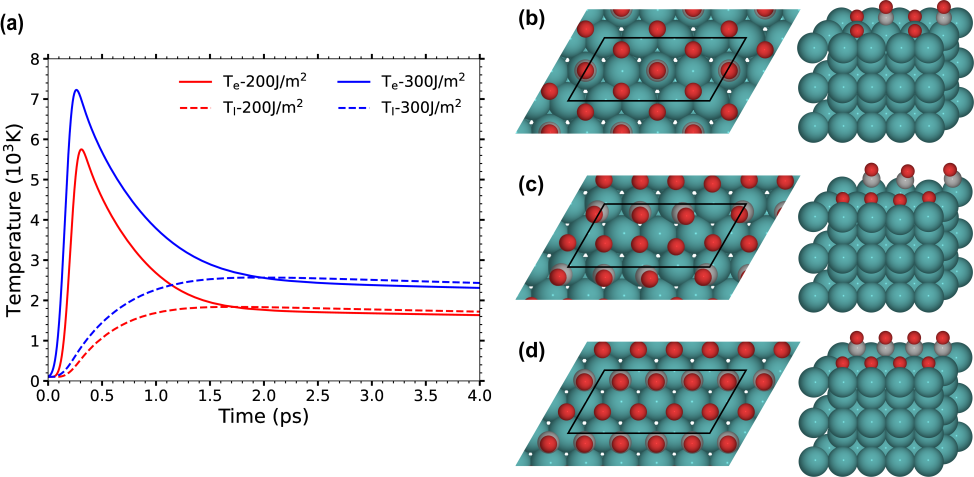}%
\end{center}
\caption{(a) 2TM-calculated electronic (blue) and lattice (red) temperatures induced by a 800~nm laser pulse with 110~fs FWHM and $F=$~200 and 300~J/m$^2$. 
(b)-(d) Top and perspective views of the energetically favored structures found in~\citep{tetenoire2021} for Ru(0001) covered by: (b) 0.5~ML O + 0.25~ML CO, (c) 0.5~ML O + 0.375~ML CO, and (d) 0.5~ML O + 0.5~ML CO. The black parallelograms depict the (4$\times$2) surface unit cell used in the AIMDEF calculations for each coverage. Color code: O atoms in red, C in gray, and Ru in blue. For clarity, the periodic images of the O and CO adsorbates in the perspective view are not shown. Images prepared with ASE~\citep{ase}.
  }\label{fig:2TM_cell}
\end{figure}

The photoinduced desorption and oxidation of CO from the (O,CO)-covered Ru(0001) surface was simulated in \citep{tetenoire2022, tetenoire2023} with the ab initio classical molecular dynamics with electronic friction method ($T_\textrm{e}, T_\textrm{l}$)-AIMDEF that allows to include the effect of both the laser-induced hot electrons and concomitant electron-excited phonons~\citep{alducin2019}. As described in detail elsewhere~\citep{alducin2019,tetenoire2022}, the electronic and ensuing phononic excitations created in the metal surface by near infrared laser pulses are described within a two-temperature model (2TM)~\citep{anisimov74} in terms of two coupled heat thermal baths. The time-dependent temperatures that are associated to the electron and phonon baths, $T_\textrm{e}(t)$ and $T_\textrm{l}(t)$, are obtained by solving the following differential equations:
\begin{eqnarray}
C_{e}\frac{\partial T_{e}}{\partial t}& =& \frac{\partial}{\partial
z}\kappa\frac{\partial T_{e}}{\partial z}-g\,(T_{e}-T_{l})+S(z,t) \, , \label{eq:TTM_e}
\\
C_{l}\frac{\partial T_{l}}{\partial t} & =& g(T_{e}-T_{l}) \, , \label{eq:TTM_l}
\end{eqnarray}
where $C_{e}$ and $C_{l}$ are the electron and phonon heat capacities, respectively, $\kappa$ is the electron thermal conductivity, $g$ is the electron-phonon coupling constant, and $S(z,t)$ is the absorbed laser power per unit volume that depends on the shape, wavelength, and fluence of the applied pulse. According to the above equations, the laser pulse is responsible of heating directly the electron system that subsequently transfers part of its energy into either the bulk electrons or the lattice phonons [first and second terms in the r.h.s. of Equation~(\ref{eq:TTM_e}), respectively]. The diameter of the laser beam, on the one hand, and the time scale of few tens of picoseconds of interest, on the other hand, justify neglecting lateral heat diffusion by electrons in Equation~(\ref{eq:TTM_e}) and heat diffusion by phonons in Equation~(\ref{eq:TTM_l})~\citep{frischkorncr06}. All the simulations performed in the present work as well as those in~\citep{tetenoire2022, tetenoire2023} correspond to irradiating the surface with the experimental pulse of ref.~\citep{bonn99}, i.e., a 800~nm Gaussian pulse of 110~fs duration. Figure~\ref{fig:2TM_cell} shows the results for $T_\textrm{e}(t)$ and $T_\textrm{l}(t)$ as obtained from 2TM for the experimental absorption fluences $F=$~200 and 300~J/m$^2$. As input parameters for the Ru(0001) surface in Equations~(\ref{eq:TTM_e}) and~(\ref{eq:TTM_l}), we use those of refs.~\citep{vazha09,scholz16,juaristiprb17,tetenoire2022}.

Next, the effect of the laser-excited electrons on each adsorbate is described through the following Langevin equation:
\begin{equation}\label{eq:langevin}
m_i\frac{d^2\mathbf{r}_i}{dt^2}=-\nabla_{\mathbf{r}_i} 
V(\mathbf{r}_1,...,\mathbf{r}_N)-\eta_{e,i}(\mathbf{r}_i)\frac{d\mathbf{r}_i}{dt} 
+\mathbf{R}_{e,i}[T_{e}(t),\eta_{e,i}(\mathbf{r}_i)] \, , 
\end{equation}
where $m_i$, $\textbf{r}_i$, and $\eta_{e,i}$ are the mass, position vector, and electronic friction coefficient of the $i^{th}$ atom conforming the set of adsorbates. The adiabatic force [first term in the r.h.s. of Equation~(\ref{eq:langevin})] depends on the position of all  atoms in the system (i.e., adsorbates and surface atoms). The electronic friction force (second term) and the electronic stochastic force (third term), which are related by the fluctuation-dissipation theorem, describe the effect of the electronic excitations and deexcitations on the adsorbate dynamics. In particular, $\mathbf{R}_{e,i}$ is modeled by a Gaussian white noise with variance, 
$\mathrm{Var}[\mathbf{R}_{e,i}(T_{e},\eta_{e,i})]=(2 k_B T_{e}(t)
\eta_{e,i}(\mathbf{r}_i))/\Delta t $,
with $k_B$ the Boltzmann constant and $\Delta t$ the time-integration step. For each atom $i$, the electronic friction coefficient $\eta_{e,i}(\mathbf{r}_i)$ is calculated with the local density friction approximation (LDFA)~\citep{juaristi08,alducinpss17}. Within this approximation, the friction coefficient is assumed to be equal to the friction
coefficient that the same atom $i$ would experience in case of moving within a homogeneous free electron gas (FEG) of
density $n_0 = n_\textrm{sur}(\textbf{r}_i)$, with $n_\textrm{sur}(\textbf{r}_i)$ being the electron density of the bare metal surface at the position $\textbf{r}_i$. As proposed by Novko \textit{et al.}~\citep{novkoprb15,novkoprb16}, an efficient method to extract on-the-fly the bare surface electron density from the self-consistent DFT electron density of the whole system (adsorbates and surface), which is calculated at each integration step in AIMDEF, consists in applying the Hirshfeld partitioning scheme~\citep{Hirshfeld1977}. Specifically, the latter is used to subtract the contribution of the adsorbates from the self-consistent electronic density in order to obtain the bare surface electron density.

In the ($T_\textrm{e}, T_\textrm{l}$)-AIMDEF simulations also the heating of the surface lattice due to the laser-induced electronic excitations is included. The latter is achieved by coupling the surface atoms 
to a Nos\'{e}-Hoover thermostat~\citep{Nose84,Hoover1985} that follows the temperature $T_\textrm{l}(t)$ obtained from 2TM. In contrast, in the $T_\textrm{e}$-AIMDEF simulations that we perform in this work all the surface atoms are kept fixed at their equilibrium positions and only the adsorbates are allowed to move as dictated by the $T_\textrm{e}$-dependent Langevin dynamics [Equation~(\ref{eq:langevin})]. These dynamics-restricted simulations are an attempt to single out the direct effect of the laser-excited electrons on the adsorbates from the effect due to energy transfer between the adsorbates and the surface atoms, which are also vibrationally excited by the electrons. 

\subsection{ General DFT Computational Settings}
The new $T_\textrm{e}$-AIMDEF simulations presented here were performed with {\sc vasp}~\citep{vasp1,vasp2} (version 5.4) and the AIMDEF module~\citep{blanco14,saalfrank14,novkoprb15,novkoprb16,novkonimb16,novkoprb17,juaristiprb17} using the same computational settings that we used in our previous ($T_\textrm{e}, T_\textrm{l}$)-AIMDEF simulations of the desorption and oxidation of CO on different covered Ru(0001) surfaces~\citep{tetenoire2022,tetenoire2023}. Figure~\ref{fig:2TM_cell} shows the supercells used to characterized the three coverages under study:
\begin{itemize}
    \item The low coverage (0.5ML~O+0.25ML~CO), in which each atop CO is surrounded by six O atoms that adsorb on the nearest hcp and fcc sites forming a honeycomb arrangement.
    \item  The intermediate coverage (0.5ML~O+0.375ML~CO), in which the O atoms adsorb at hcp sites forming a p(1$\times$2) structure, while the CO molecules occupy the empty space left between the O arrays.
    \item The high coverage (0.5ML~O+0.5ML~CO), in which both the O and CO adsorb on hcp sites forming two inserted p(1$\times$2) structures.
\end{itemize}
As seen in the figure, the three coverages are modeled with the same supercell that consists of a (4$\times$2) surface unit cell and a vector length along the surface normal of 30.22~{\AA}. Within this supercell, each covered Ru(0001) surface is described by five layers of Ru atoms and the corresponding (O,CO) overlayer. The Ru topmost layer and the bottom of the nearest periodic Ru slab are separated by about $19$~{\AA} of vacuum. The employed (4$\times$2) surface cell contains various adsorbates and, hence, it will provide a reasonable description of the interadsorbate interactions and their effect in the adsorbate dynamics, which become important at sufficiently large coverages~\citep{denzler03,xin15,sung16,juaristiprb17,alducin2019, serrano2021, lidner2023}. Let us remark that the low and intermediate coverages have been found in experiments~\citep{Kostov1992Nov}, while the high coverage is predicted to be stable by DFT~\citep{tetenoire2021} but has not been experimentally observed.

In the $T_\textrm{e}$-AIMDEF simulations, the adiabatic forces are calculated with non spin-polarized DFT using the van der Waals exchange-correlation functional proposed by~\citep{Dion2004} and the same computational parameters that were used in our previous studies on the energetics~\citep{tetenoire2021} and ($T_\textrm{e}, T_\textrm{l}$)-AIMDEF dynamics of the O+CO-Ru(0001) system~\citep{tetenoire2022,tetenoire2023}. Specifically, the electronic ground state energy is determined at each integration step within a precision of $10^{-6}$~eV. Integration in the Brillouin zone is performed using a $\Gamma$-centered 3$\times$6$\times$1  Monkhorst-Pack grid of special \textbf{k} points~\citep{Monkhorst1976} and the Methfessel and Paxton scheme of first order with a broadening of 0.1~eV~\citep{Methfessel1989}. The Kohn-Sham orbitals are expanded in a plane-wave basis set with an energy cutoff of $400$~eV. The projector augmented wave (PAW) method~\citep{Blochl1994Dec} that is implemented in VASP~\citep{Kresse1999} is used to describe the electron-core interaction. Integration of the Langevin equation is performed with the Beeman method implemented in our AIMDEF module~\citep{blanco14}. Each trajectory starts with the adsorbates at rest at their equilibrium position and is propagated up to 4~ps using a time step of 1~fs. For each coverage and absorbed fluence we run 100 trajectories.

\subsection{Calculation of observables}
Following \citep{tetenoire2022,tetenoire2023}, a CO molecule is counted as desorbed if its center of mass height reaches the distance $Z_\textrm{cm}=$~6.5~{\AA} from the Ru(0001) topmost layer with positive momentum along the surface normal ($P_\textrm{z} > 0$). After analyzing all the trajectories, the  CO oxidation (i.e., the O+CO recombinative desorption as CO$_2$) and CO desorption probabilities per CO molecule are calculated for each coverage as
\begin{equation}
    P_\textrm{des}(A)=\frac{N_\textrm{des}(A)}{N_\textrm{t} N_\textrm{CO}}
\end{equation}
with $N_\textrm{des}(A)$ the number of the desorbing molecules under consideration (i.e., $A$ stands for CO or CO$_2$), $N_\textrm{t}$ the total number of trajectories, and $N_\textrm{CO}$ the number of CO molecules in the simulation cell (2, 3, and 4, respectively, for low, intermediate, and high coverages). 

The mean total kinetic energy $\langle E_\textrm{kin}\rangle (t)$ and mean center-of-mass kinetic energy $\langle E_\textrm{cm}\rangle (t)$ per adsorbate type are calculated at each instant $t$ as
\begin{equation}
    \langle E_\textrm{kin (cm)}\rangle (t)=\sum_{i=1}^{N_\textrm{t}} \sum_{j=1}^{N_a}\frac{E_\textrm{kin (cm)}^j(t)}{N_\textrm{t} N_a}
\end{equation}
where $N_a$ is the total number of the specific species under consideration (e.g., nondesorbing CO molecules that remain adsorbed on the surface at the end of the simulation, CO molecules that desorb, nondesorbing O adatoms...) and $E_\textrm{kin (cm)}^j$ is the kinetic (center-of-mass) energy of adsorbate $j$ at instant $t$.

\section{Results and Discussion\label{sec:results}}
The CO desorption and CO oxidation probabilities obtained from the $T_\textrm{e}$-AIMDEF and  ($T_\textrm{e}, T_\textrm{l}$)-AIMDEF simulations at the same absorbed fluence of 200~J/m$^2$ are compared for each coverage in Table~\ref{tab:table1}. The CO desorption probabilities in the intermediate and high coverages are reduced by a factor 33.8 and 34.5, respectively, when only the direct effect of the excited electrons are included ($T_\textrm{e}$-AIMDEF). Assuming that a similar factor of $\sim$34 stands for the low coverage, we consider that the predicted desorption probability of $\sim 0.5\%$ is compatible with the lack of CO desorption events we obtain within our limited statistics. As found in the ($T_\textrm{e}, T_\textrm{l}$)-AIMDEF simulations~\citep{tetenoire2023}, the $P_\textrm{des}(\textrm{CO}$) values correlate well with the CO desorption barriers calculated with DFT-vdW for each coverage~\citep{tetenoire2021}. That is, the number of CO desorption events increases as the barrier decreases. Let us remark that the drastic reduction we obtain in the $T_\textrm{e}$-AIMDEF desorption probabilities aligns with the two-pulse correlation measurements suggesting that the photoinduced desorption of CO on the O+CO-Ru(0001) surface is a phonon-dominated process~\citep{bonn99}. Interestingly, this feature, the importance of the excited phonons in the photodesorption of CO, is not exclusive of the (O,CO)-covered surface, as it has been observed in diverse experiments in which Ru(0001) is covered with CO~\citep{funk00} and in molecular dynamics calculations motivated by those experiments~\citep{scholz16}, which included the effect of $T_e(t)$ and $T_l(t)$ following the model by~\citep{loncaricprb16}. 

In respect of the CO oxidation process, there are no events in the case of the $T_\textrm{e}$-AIMDEF simulations. Nevertheless, the statistics is insufficient to exclude that the laser-excited electrons are the dominant driving mechanism, as proposed in~\citep{bonn99}. The analysis of the kinetic energy and displacements below will show however that there are distinct features in the ($T_\textrm{e}, T_\textrm{l}$)-AIMDEF adsorbate dynamics as compared to the $T_\textrm{e}$-AIMDEF adsorbate dynamics suggesting that not only electrons but also the highly excited phonons are contributing to the oxidation process, similarly to what was obtained for the laser-induced desorption of CO from Pd(111)~\citep{alducin2019}. 

In order to confirm the above idea and gain further insights into the role of the excited electrons and phonons we also calculated for illustrative purposes an additional set of 100 $T_\textrm{e}$-AIMDEF trajectories assuming a extreme absorption fluence $F=$~300~J/m$^2$ for one of the covered surfaces, namely, the high coverage. As shown in Figure~\ref{fig:2TM_cell}(a), the maximum of the electronic temperature for the new fluence is about 1600~K higher than for $F=$~200~J/m$^2$. After reaching the maximum, a difference of about 800~K is still maintained during the rest of the integration time used in our calculations. The purpose of these new simulations is to increase the energy provided to the adsorbates but excluding effects due to the lattice distortions inherent to phonon excitations. The results in Table~\ref{tab:table1} show that $P_\textrm{des}(\textrm{CO}$) increases from 1\% to 15.25\% because of the fluence. The latter value is still about a factor 2 smaller than in the ($T_\textrm{e}, T_\textrm{l}$)-AIMDEF simulations for $F=200$~J/m$^2$. Lastly, neither at this high fluence there are oxidation events, although the analysis of the adsorbate displacements below will show that in a few cases the adsorbates can eventually abandon their adsorption well. 

\begin{table*}
\caption{\label{tab:table1} $T_\textrm{e}$-AIMDEF CO desorption probability $P_\textrm{des}$(CO) and CO oxidation probability $P_\textrm{des}$(CO$_2$) calculated for the low, intermediate, and high (O,CO)-Ru(0001) coverages at an absorbed fluence $F=$~200~J/m$^2$. For the high coverage also results at $F=$~300~J/m$^2$ are shown (last row) . For comparison, the probabilities obtained from  ($T_\textrm{e}$,$T_\textrm{l}$)-AIMDEF simulations with $F=$~200~J/m$^2$ in ref~\citep{tetenoire2023} are reproduced within parenthesis. Activation energies (in eV) for CO desorption $E_\textrm{TS}^\textrm{CO}$ and CO oxidation $E_\textrm{TS}^{\textrm{CO}_2}$ are from ref~\citep{tetenoire2021}.} %
\begin{tabular}{cccccc}
\hline \vspace{0.1cm}
Coverage & Fluence (J/m$^2$)&$P_\textrm{des}$(CO) (\%) & $P_\textrm{des}$(CO$_2$) (\%)  &$E_\textrm{TS}^\textrm{CO}$ (eV) & $E_\textrm{TS}^{\textrm{CO}_2}$ (eV)\\ \hline 
0.5ML~O+0.250ML~CO &200&  0.00 (18.25) &  0.00 (0.50) &1.57 & 1.19\\
0.5ML~O+0.375ML~CO & 200&  1.33 (45.06) &  0.00 (0.67) & 0.58,0.73& 0.80\\
0.5ML~O+0.5ML~CO& 200 &  1.00 (34.53) &  0.00 (1.26) & 0.88 & 2.01\\
0.5ML~O+0.5ML~CO& 300 &  15.25 ( -- ) &  0.00 ( -- ) & 0.88 & 2.01\\

\end{tabular}
\end{table*}

\subsection{Kinetic Energy Gain}

\begin{figure}[]
\begin{center}
\includegraphics[width=1.0\textwidth]{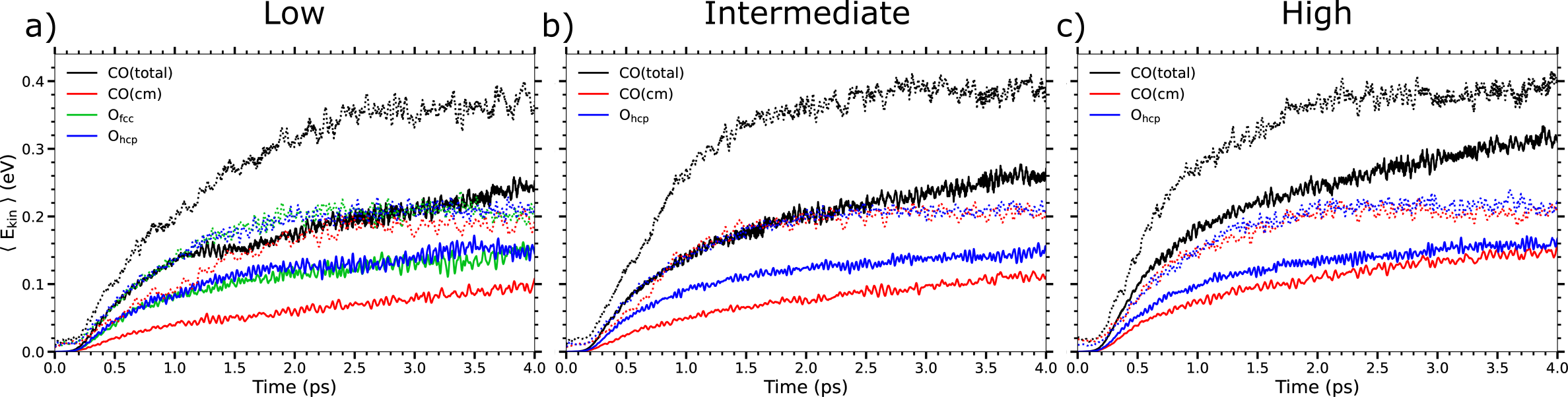}%
\end{center}
\caption{Time evolution of the adsorbate mean kinetic energy in the low (a), intermediate (b), and high  (c) coverages as obtained in $T_\textrm{e}$-AIMDEF (solid lines) and ($T_\textrm{e}$,$T_\textrm{l}$)-AIMDEF (dotted lines) for an absorbed laser fluence $F=$~200~J/m$^2$. Shown for each coverage are: the mean total kinetic energy $\langle E_\textrm{kin}\rangle$ (black) and mean center-of-mass kinetic energy $\langle E_\textrm{cm}\rangle$ (red) of nondesorbing CO and the mean total kinetic energy of nondesorbing O$_\textrm{hcp}$ (blue) and O$_\textrm{fcc}$ (green).}\label{fig:ekin}
\end{figure}

The time evolution of the mean kinetic energy of the adsorbates along the $T_\textrm{e}$-AIMDEF (thick solid lines) and ($T_\textrm{e}$,$T_\textrm{l}$)-AIMDEF dynamics (dotted lines) is compared in Figure~\ref{fig:ekin} for each adsorbate type and each coverage. In both simulations the absorbed laser fluence is $F=$~200~J/m$^2$. For simplicity, only the results of the nondesorbing species, i.e., the adsorbates that remain on the surface at the end of our simulations, will be discussed. A detailed analysis of the kinetic energy gained by the desorbed CO in the ($T_\textrm{e}$,$T_\textrm{l}$)-AIMDEF simulations can be found elsewhere~\citep{tetenoire2023}. A common observation in Figure~\ref{fig:ekin} is that irrespective of the coverage the adsorbates gain less kinetic energy in the $T_\textrm{e}$-AIMDEF simulations than in ($T_\textrm{e}$,$T_\textrm{l}$)-AIMDEF. 
There exist some interesting features worth mentioning. As discussed in ~\citep{tetenoire2023} in the case of ($T_\textrm{e}$,$T_\textrm{l}$)-AIMDEF simulations a quasithermalized state was obtained at the end of the simulations, and even more rapidly for the intermediate and high coverages. This is shown by the fact that the average total kinetic energy of the CO molecules is twice the average kinetic energy of their center of mass, and that this coincides, roughly, with the average kinetic energy of the O atoms. This is what is expected when there exists equipartition of the energy among the different degrees of freedom. This is clearly not the case in the $T_\textrm{e}$-AIMDEF simulations. For instance, we observe that for all coverages the average kinetic energy of the atomic O is larger than the average kinetic energy of the center of mass of the CO molecules. This can be rationalized by the fact that the O atoms are more strongly bound than the CO molecules and therefore their coupling to the electronic system is stronger. Note, finally, that even though this statement is generally true for all the coverages, in the high coverage case the difference between these two energies is much smaller and that tends to disappear at the end of the simulation time. This may be due to an increased importance of interadsorbate energy exchange that favors the thermalization of the system when the concentration of adsorbates at the surface is larger.

\begin{figure}[]
\begin{center}
\includegraphics[width=0.5\columnwidth]{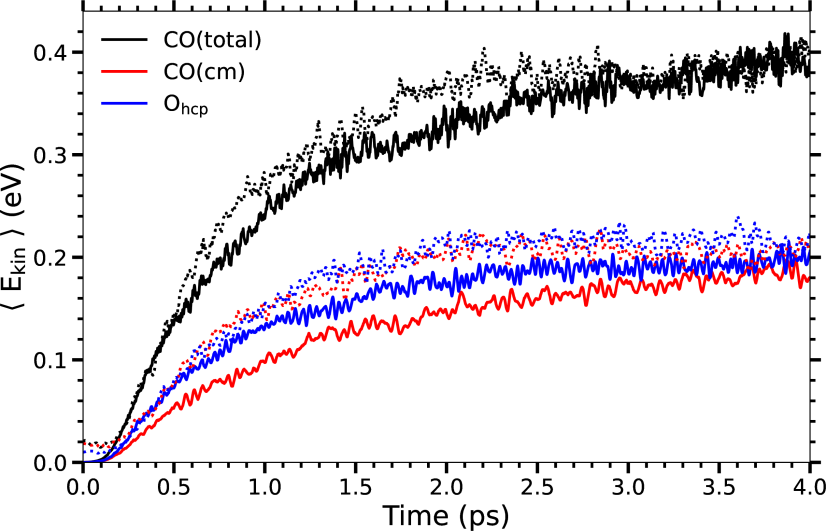}%
\end{center}
\caption{Same as Figure~\ref{fig:ekin} for the high coverage. The results from $T_\textrm{e}$-AIMDEF at $F=$~300~J/m$^2$ (solid lines) are compared to those from ($T_\textrm{e}$,$T_\textrm{l}$)-AIMDEF at $F=$~200~J/m$^2$ (dotted lines). }\label{fig:ekin_high}
\end{figure}

Nevertheless, note that the energy gain when only electronic excitations are considered is roughly one half of the energy gain when both electronic and phononic excitations are taken into account. It is difficult to rationalize reduction factors larger than 30 in the CO desorption probabilities such as those presented in Table~\ref{tab:table1} in terms of this reduction in the energy gain. This suggests that not only the increased energy gain but also other effects are playing a role in the increased CO desorption and oxidation probabilities when phononic excitations are considered. In order to strengthen this point further, in Figure~\ref{fig:ekin_high} we show the results for the kinetic energy gain of the adsorbates in $T_\textrm{e}$-AIMDEF simulations in the high coverage case for a larger fluence, namely $F=300$~J/m$^2$. In this case, albeit a slightly different time dependence, the energy gain is very similar to that obtained with $F=200$~J/m$^2$ in the ($T_\textrm{e}$,$T_\textrm{l}$)-AIMDEF simulations. Nevertheless, even with similar energy gains desorption and oxidation probabilities are, as shown in Table~\ref{tab:table1}, much lower when only electronic excitations are accounted for. This constitutes a definitive proof of the fact that the role played by phononic excitations in the desorption and oxidation probabilities is not limited to being a energy source channel. This important point is further analyzed in the next subsection.

\subsection{Adsorbate Displacements}

Evidence of the important role of the excited phonons in the photoinduced reactions on (O+CO-Ru(0001) is provided by comparing the in-plane displacement of the adsorbed species between both types of calculations, $T_\textrm{e}$-AIMDEF and ($T_\textrm{e},T_\textrm{l}$)-AIMDEF. As in the previous section, only the diffusion of the nondesorbing species will be discussed. 

\begin{figure}[]
\begin{center}
\includegraphics[width=1.0\columnwidth]{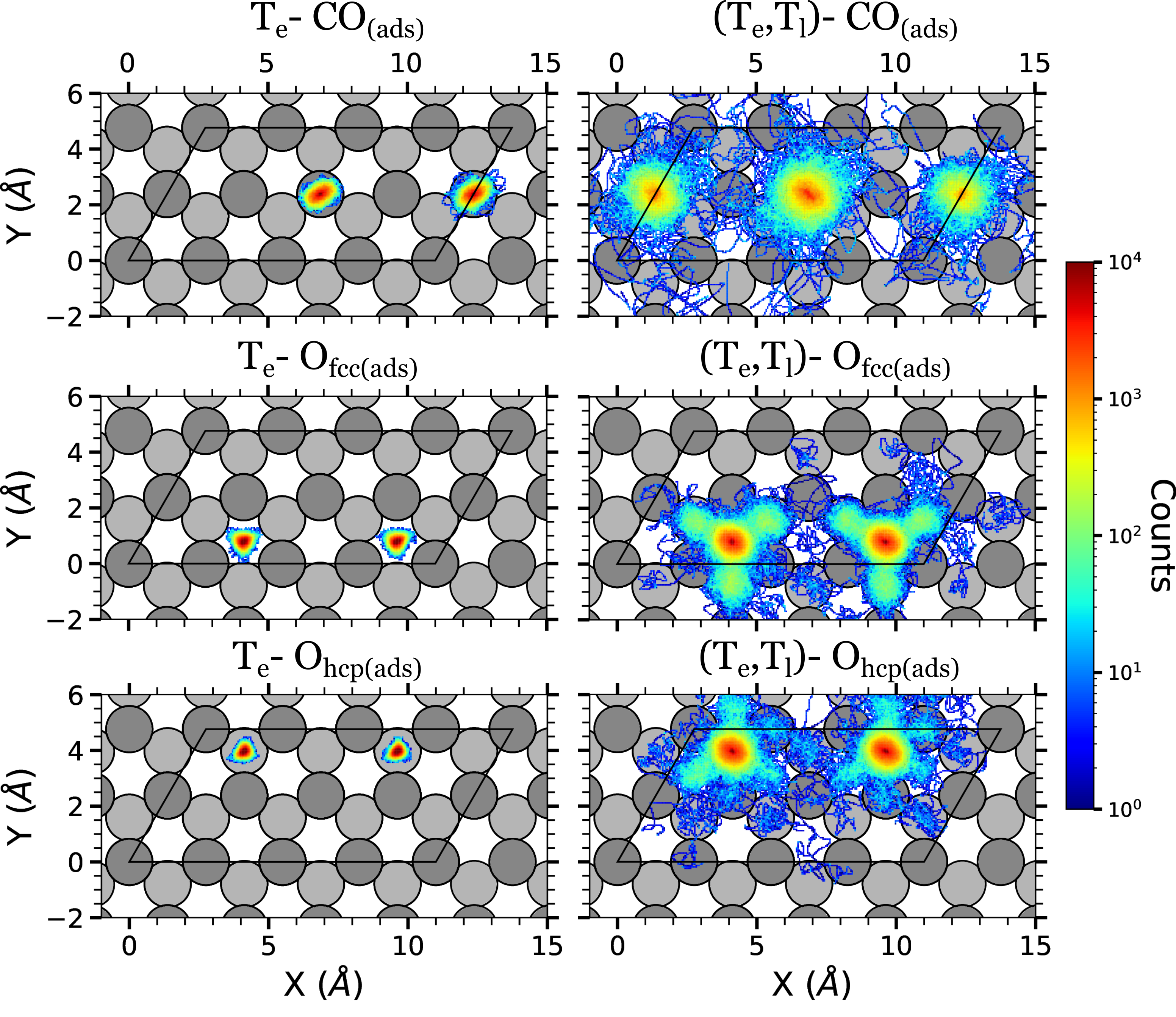}%
\end{center}
\caption{Density plots of the $(x,y)$ positions over the surface of the adsorbates at the low surface coverage in the AIMDEF simulations. Only the positions of the adsorbates that remain adsorbed on the surface at the end of the simulation are shown. Left (right) column shows the results of the  $T_\textrm{e}$-AIMDEF [($T_\textrm{e}$,$T_\textrm{l}$)-AIMDEF] simulations. Top panels correspond to the center of mass of CO molecules, middle panels to the position of O atoms initially on fcc sites, and bottom panels to the position of O atoms initially on hcp sites. The black line encloses the simulation cell. For clarity, the position of the adsorbates is shown in an extended coordinate representation.}
  \label{fig:xy_low}
\end{figure}

\begin{figure}[]
\begin{center}
\includegraphics[width=1.0\columnwidth]{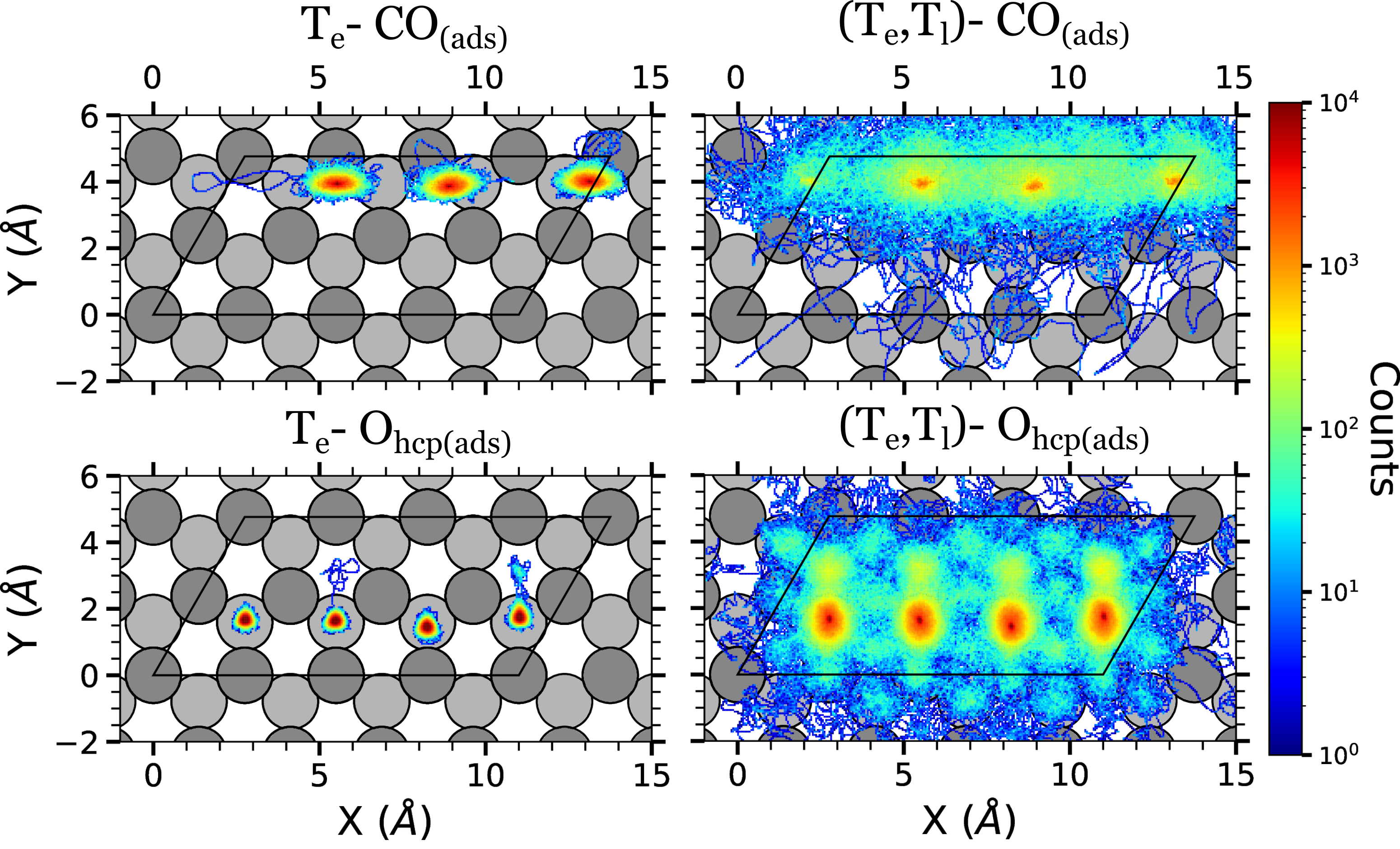}
\end{center}
\caption{ Density plots of the $(x,y)$ positions over the surface of the adsorbates at the intermediate surface coverage in the AIMDEF simulations. Only the positions of the adsorbates that remain adsorbed on the surface at the end of the simulation are shown. Left (right) column shows the results of the  $T_\textrm{e}$-AIMDEF [($T_\textrm{e}$,$T_\textrm{l}$)-AIMDEF] simulations. Top panels correspond to the center of mass of CO molecules and bottom panels to the position of O atoms. The black line encloses the simulation cell. For clarity, the position of the adsorbates is shown in an extended coordinate representation.}\label{fig:xy_intermediate}
\end{figure}

\begin{figure}[]
\begin{center}
\includegraphics[width=1.0\columnwidth]{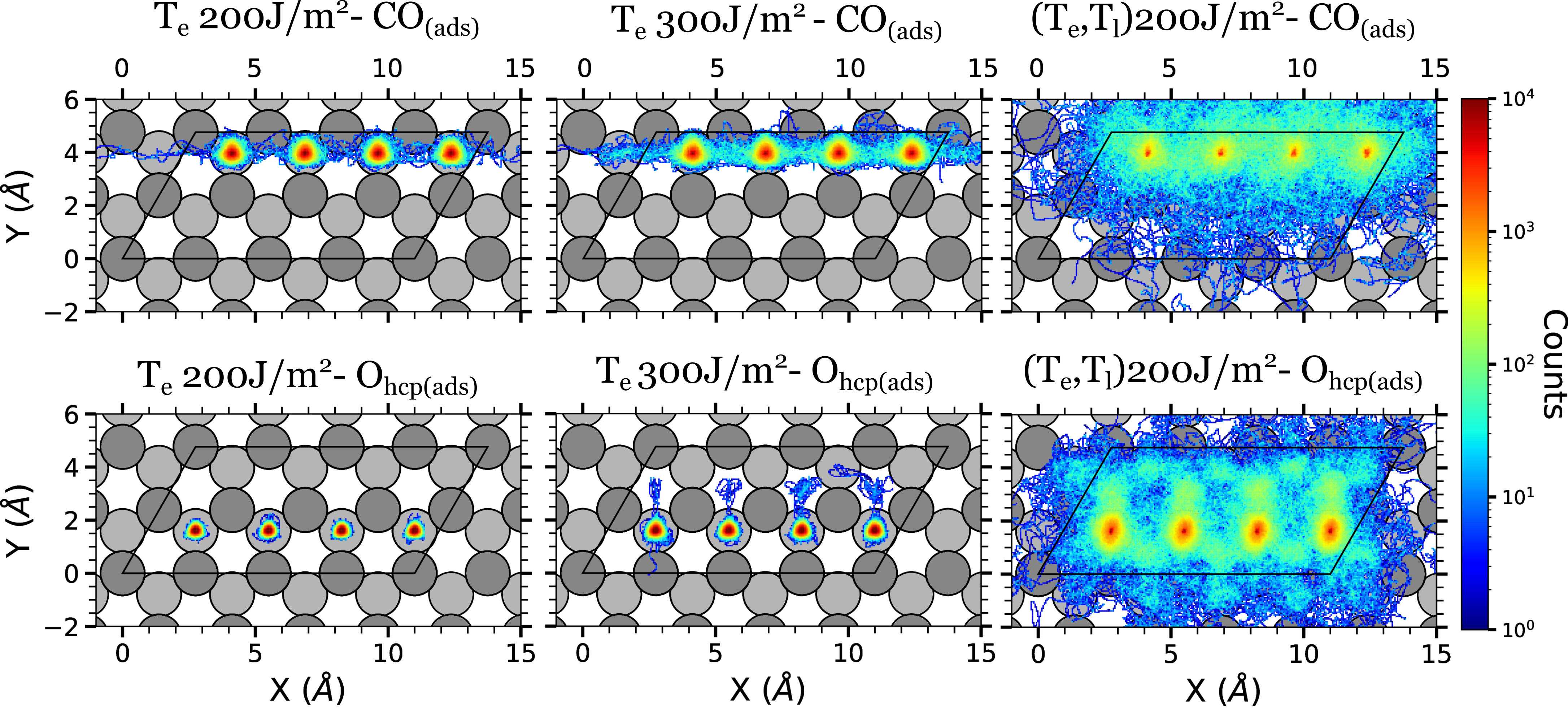}
\end{center}
\caption{Density plots of the $(x,y)$ positions over the surface of the adsorbates at the high surface coverage in the AIMDEF simulations. Only the positions of the adsorbates that remain adsorbed on the surface at the end of the simulation are shown. Left, center, and right columns show the results of the $T_\textrm{e}$-AIMDEF with $F=200$~J/m$^{2}$, $T_\textrm{e}$-AIMDEF with $F=300$~J/m$^{2}$, and ($T_\textrm{e}$,$T_\textrm{l}$)-AIMDEF with $F=200$~J/m$^{2}$ simulations, respectively. Top panels correspond to the center of mass of CO molecules and bottom panels to the position of O atoms. The black line encloses the simulation cell. For clarity, the position of the adsorbates is shown in an extended coordinate representation.}\label{fig:xy_high}
\end{figure}

The displacement data is presented in terms of colored density plots, which correspond to two-dimensional histograms of the adsorbates ($x$,$y$) positions over the surface (Figures~\ref{fig:xy_low}, \ref{fig:xy_intermediate}, and \ref{fig:xy_high}). For each kind of adsorbate and simulation type, each density plot is constructed using the in-plane positions along the whole trajectory (4~ps, \textit{i.e.,} 4000 steps) of all the adsorbates of that kind and of all the simulated trajectories. Thus, the density color code gives qualitatively an idea of the amount of time the adsorbates have spent in a given position (higher densities will correspond to longer times). Let us also remark that in each plot the atoms are allowed to go out of the unit cell (enclosed by a black solid line in the figures). The reason is that we are using a extended coordinate representation in order to show the continuous path followed by the adsorbates. 

%

The adsorbate displacements in the low coverage case are plotted in Figure~\ref{fig:xy_low}. In the $T_\textrm{e}$-AIMDEF simulations (left panels of Figure~\ref{fig:xy_low}), the O adatoms stay on (or very close to) their respective adsorption sites. Something similar is observed for the CO molecules. They remain in top/near-top sites, showing no preference to move neither towards fcc sites nor towards hcp sites. The CO molecules explore an ellipse centered on the top position with  a long axis of $\sim$1.3~{\AA} and a short axis of $\sim$1~{\AA}. These short displacements are clearly insufficient for CO oxidation to occur because the CO and O adsorbates cannot get close enough to recombine. 

The in-plane mobility of all the adsorbates increases much when the Ru lattice excitation is incorporated with the ($T_\textrm{e}$,$T_\textrm{l}$)-AIMDEF simulations (right panels of Figure~\ref{fig:xy_low}). The O adatoms can now abandon their initial adsorption site and cross the surrounding bridge sites. In particular, O atoms initially located at the fcc sites (O$_\textrm{fcc}$) show a tendency to move to the nearest hcp sites, whereas the ones initially located at the hcp sites (O$_\textrm{hcp}$) show a tendency to move to the closest fcc sites. We also observe that O$_\textrm{hcp}$ shows a slightly smaller mobility than O$_\textrm{fcc}$. This is consistent with the larger adsorption energy of the former (5.62~eV) as compared to the latter (4.95~eV)~\citep{tetenoire2023}. Although less probable, note that there also exist events in which the O atoms move beyond the nearest neighbor adsorption sites. The mobility of the CO molecules is also much increased in the ($T_\textrm{e}$,$T_\textrm{l}$)-AIMDEF simulations. Now they explore a circle of radius $\sim$2.3~{\AA} centered at their equilibrium position. In some cases, the excited CO may even move beyond their first neighboring site. 

Figure~\ref{fig:xy_intermediate} shows that the adsorbate mobility is drastically reduced also in the intermediate coverage when the effect of the hot Ru lattice is not included. In the $T_\textrm{e}$-AIMDEF simulations (left column of Figure~\ref{fig:xy_intermediate}), the CO molecules basically move within an ellipse centered at their corresponding  adsorption site as in the low coverage case, although the explored area is larger (long and short axis lengths of $\sim$2.5~{\AA} and $\sim$1.3~{\AA}, respectively). We also observe a few cases in which the CO diffuses either along the $x$ direction or to a nearest top site. The O adatoms mostly remain in their hcp adsorption sites, although there exist a few events in which O diffuses towards the nearest fcc site that is located farther from the other adatoms. Furthermore, these diffusing atoms are the ones that have not a CO molecule adsorbed on the near-top site that is located above them in the density plot. We checked that the displacement of the second left O occurs once the nearest CO above it desorbs. 

The mobility of both kind of adsorbates increases considerably in the ($T_\textrm{e}$,$T_\textrm{l}$)-AIMDEF simulations (right column of Figure~\ref{fig:xy_intermediate}). In fact, the difference respect to the $T_\textrm{e}$-AIMDEF simulations is even more pronounced than in the low coverage because in the intermediate coverage basically every spot in the simulation cell is at some instant occupied by either O or CO. Still, we observe that in the case of CO molecules, the O row acts as a barrier that prohibits the CO molecules to access the lower part of the simulation cell, except for very few rare events. In contrast, the O  adatoms can move all over the cell albeit it is less probable to find them at top sites than on hcp or fcc sites. These features provide, in a qualitative manner, indirect information on the properties of the potential energy surface and were already discussed in detail elsewhere~\citep{tetenoire2023}. 

The in-plane displacement of the adsorbates in the high coverage are shown in Figure~\ref{fig:xy_high}. Recall that in this case, the $T_\textrm{e}$-AIMDEF simulations were performed for two different fluences, namely, $F=$~200 and 300~J/m$^{2}$. In $T_\textrm{e}$-AIMDEF simulations with a laser fluence of 200~J/m$^{2}$ (left column of Figure~\ref{fig:xy_high}), the in-plane motion of the CO molecules is mostly restricted to a circle of radius $\sim$1~{\AA}. Still, we observe some events that involve lateral displacement of the CO molecules from one hcp site to another along the row in which they are located. However, since O atoms hardly moves away from their corresponding adsorption sites, no CO oxidation event is expected to take place under these conditions. 

As in the intermediate coverage, all adsorbates become extremely mobile when we also include the effect of the excited phonons [right column of Figure~\ref{fig:xy_high}, ($T_\textrm{e}$,$T_\textrm{l}$)-AIMDEF simulations]. In fact the pattern of the density plots for both coverages, which share the same $p(1\times2)$ arrangement of the O adatoms, is very similar. Specifically, the CO molecules move predominantly along the row in which they are adsorbed, while the O adatoms end moving all over the surface.

The comparison of the O and CO displacements in the $T_\textrm{e}$-AIMDEF simulations with $F=$~300~J/m$^{2}$ (middle column of Figure~\ref{fig:xy_high}) and in the ($T_\textrm{e}$,$T_\textrm{l}$)-AIMDEF simulations with $F=$~200~J/m$^{2}$ (right column of Figure~\ref{fig:xy_high}) is probably the most clear evidence of the importance that the hot phonons created indirectly by the laser pulse has on the reaction dynamics. As shown in the previous section, the energy gained by both O and CO is very similar in the two simulations (see Figure~\ref{fig:ekin_high}). In spite of it, we show here that the mobility of the adsorbates is still very limited in the high fluence $T_\textrm{e}$-AIMDEF simulations as compared to the displacements obtained in the ($T_\textrm{e}$,$T_\textrm{l}$)-AIMDEF simulations at lower laser fluence. For example, the in-plane motion of the CO molecules is mostly restricted to a circle of radius $\sim$1.2~{\AA} in the former, while it basically occupies the whole surface in the latter. Nonetheless, it is also worth noticing that compared to $T_\textrm{e}$-AIMDEF with $F=$~200~J/m$^{2}$, the lateral displacement along the $y$-axis in between the two nearest rows of Ru atoms is clearly much more probable in the high fluence simulations. In the case of the O adatoms, even if they remain mostly at their adsorption site, we observe some events in which they go through bridge sites, toward the nearest fcc sites. Clearly, the adsorbates have an increased mobility when the laser fluence is increased that could allow them to eventually recombine, even if there is no motion of the surface, but with a much smaller probability.

All in all, the present analysis shows that, for all coverages, the inclusion of lattice motion and phononic excitations increase the mobility of the adsorbates and allow them to explore larger regions of the configurational space. Therefore, though the low statistics does not allow us to categorically establish whether electronic or phononic excitations govern the CO oxidation process, these results strongly suggest that the role of phononic excitations cannot be neglected.

\begin{figure}[]
\begin{center}
\includegraphics[width=1.0\columnwidth]{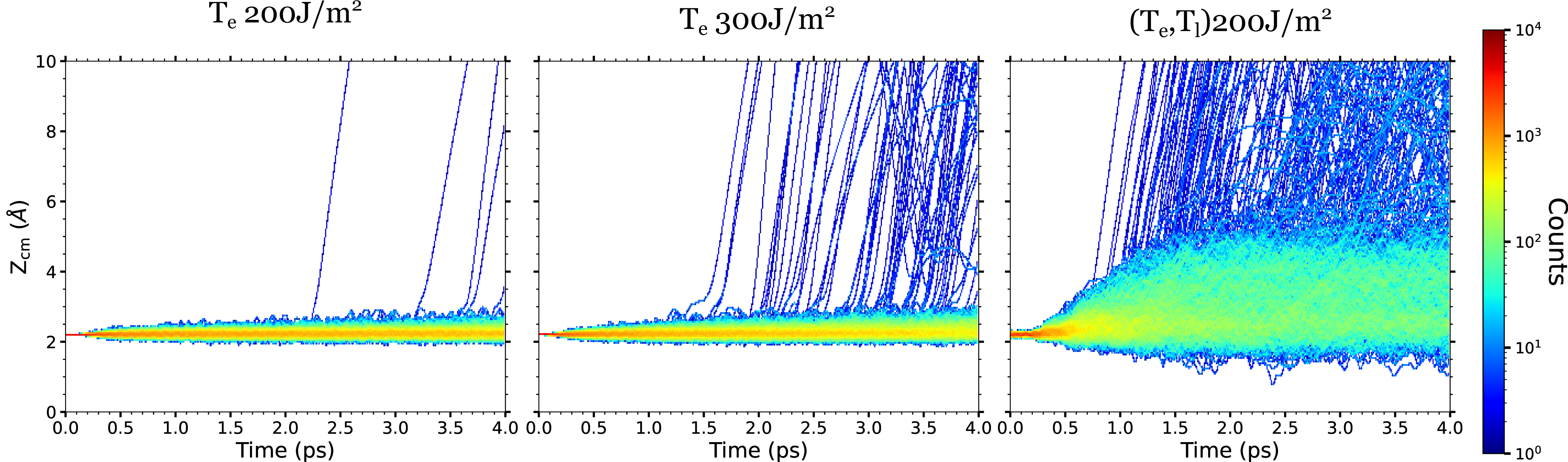}
\end{center}
\caption{Density plots of the CO center-of-mass height $Z_\textrm{CM}$ respect to the surface topmost layer obtained for the high surface coverage in the AIMDEF simulations. Left, center, and right columns show the results of the $T_\textrm{e}$-AIMDEF with $F=200$~J/m$^{2}$, $T_\textrm{e}$-AIMDEF with $F=300$~J/m$^{2}$, and ($T_\textrm{e}$,$T_\textrm{l}$)-AIMDEF with $F=200$~J/m$^{2}$ simulations, respectively.}\label{fig:zcm_high}
\end{figure}

Regarding the CO desorption process, it is also interesting to compare the displacements along the surface normal with and without including the effect of the phonon excitations. Similar to the in-plane displacements, the CO mobility along the $z$-axis is higher in ($T_\textrm{e}$,$T_\textrm{l}$)-AIMDEF than in $T_\textrm{e}$-AIMDEF for each coverage. As an example, we show in Fig.~\ref{fig:zcm_high} the time evolution of the CO center of mass height $Z_\textrm{CM}$ for the high coverage. Comparing the results calculated for the same absorbed fluence ($F=200$~J/m$^{2}$), we observe that the $Z_\textrm{CM}$ displacements of the nondesorbing CO increase from about 0.5--1~{\AA} in $T_\textrm{e}$-AIMDEF to 2--3~{\AA} in ($T_\textrm{e}$,$T_\textrm{l}$)-AIMDEF. Increasing the fluence in the $T_\textrm{e}$-AIMDEF simulations also implies an increase in the $Z_\textrm{CM}$ displacements and, importantly, in the number of desorption events, but still significantly smaller than in the ($T_\textrm{e}$,$T_\textrm{l}$)-AIMDEF simulations at 200~J/m$^{2}$. 

\section{Conclusions}

The photoinduced desorption and oxidation of CO coadsorbed with O on Ru(0001) has been simulated with ab initio molecular dynamics with electronic friction that include the effect of the laser-induced hot electrons but neglects that of the phonon excitations ($T_\textrm{e}$-AIMDEF). Comparison of these new results with those we obtained previously with simulations that incorporated in the adsorbate dynamics both the effect of the hot electrons and hot phonons [($T_\textrm{e}$,$T_\textrm{l}$)-AIMDEF)] allows us to discern the role of electrons and phonons in the oxidation and desorption of CO from the covered surface. The probability of both reactions are drastically reduced when only the coupling to electrons is included. As suggested by two pulse correlation experiments in this system, CO desorption is dominated by the transient high temperature  that is indirectly created by the laser pulse. Unfortunately, the statistics for CO oxidation is insufficient to determine the relative importance of the electronic and phononic mechanisms. Nonetheless, the comparative analysis of various dynamical properties such as the adsorbate kinetic energy and adsorbate displacements indicates that energy exchange with the hot lattice and the associated strong surface distortions are important ingredients to understand the CO oxidation reaction. This conclusion is supported by $T_\textrm{e}$-AIMDEF simulations performed at a high laser fluence. The kinetic energy gain is similar to that obtained in ($T_\textrm{e}$,$T_\textrm{l}$)-AIMDEF at a lower fluence but the adsorbate displacements are still insufficient to facilitate recombination. 

\section*{Conflict of Interest Statement}

The authors declare that the research was conducted in the absence of any commercial or financial relationships that could be construed as a potential conflict of interest.

\section*{Author Contributions}
Authors contributed equally to the development of this research project.

\section*{Funding}
The authors acknowledge financial support by the Gobierno Vasco-UPV/EHU
Project No.  IT1569-22 and by the Spanish MCIN/AEI/10.13039/501100011033 [Grant No. PID2019-107396GB-I00]. This research was conducted in the scope of the Transnational Common Laboratory (LTC) “QuantumChemPhys – Theoretical Chemistry and Physics at the Quantum Scale”. 

\section*{Acknowledgments}
Computational resources were provided by the DIPC computing center.

 
\section*{Data Availability Statement}
The data supporting the conclusions of this study are included in the manuscript, further inquiries by any qualified researcher can be directed to the corresponding author. 

\bibliographystyle{Frontiers-Harvard} %
\bibliography{ref-frontiers}


   



\end{document}